# Asymmetric jet cuts in dijet measurements at ZEUS

Sabine W Lammers
University of Wisconsin
sabine.lammers@desy.de

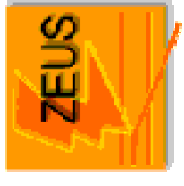

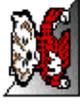

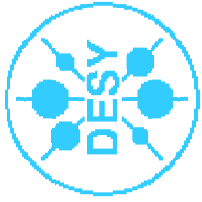

## Outline

Dijets in e–p collisions
Infrared Sensitivity at NLO
Study with DISENT
Impact on ZEUS Measurements

# Dijets in e–p collisions

Leading Order QCD Dijet Diagrams:

Boson–Gluon Fusion

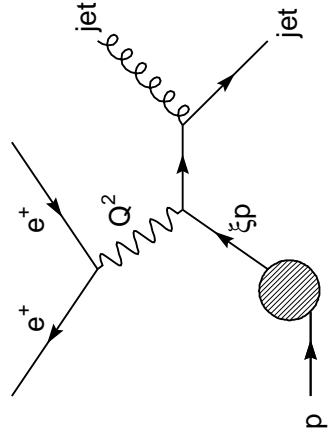

$\sigma^{2+1} \sim \hat{\sigma}_{BGF} \cdot g(x, Q^2)$

QCD Compton

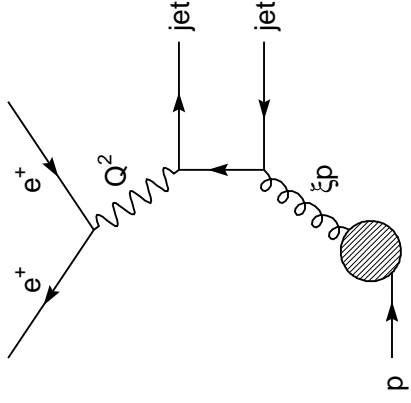

$\sigma^{2+1} \sim \hat{\sigma}_{QCDC} \cdot q(x, Q^2)$

$E_p = 920$ GeV
$E_{e+} = 27.5$ GeV
318 GeV center-of-mass energy

$\xi = x\left(1 + \dfrac{M_{jj}^2}{Q^2}\right)$

Momentum fraction of incident parton

BGF (QCDC) contribution directly proportional to gluon (quark) density in the proton.

# Asymmetric Jet Cuts I: Infrared Sensitivity

NLO calculations for BGF process receive contributions from LO Born cross sections and $O(\alpha_s^2)$ real (positive) and virtual (negative) corrections.

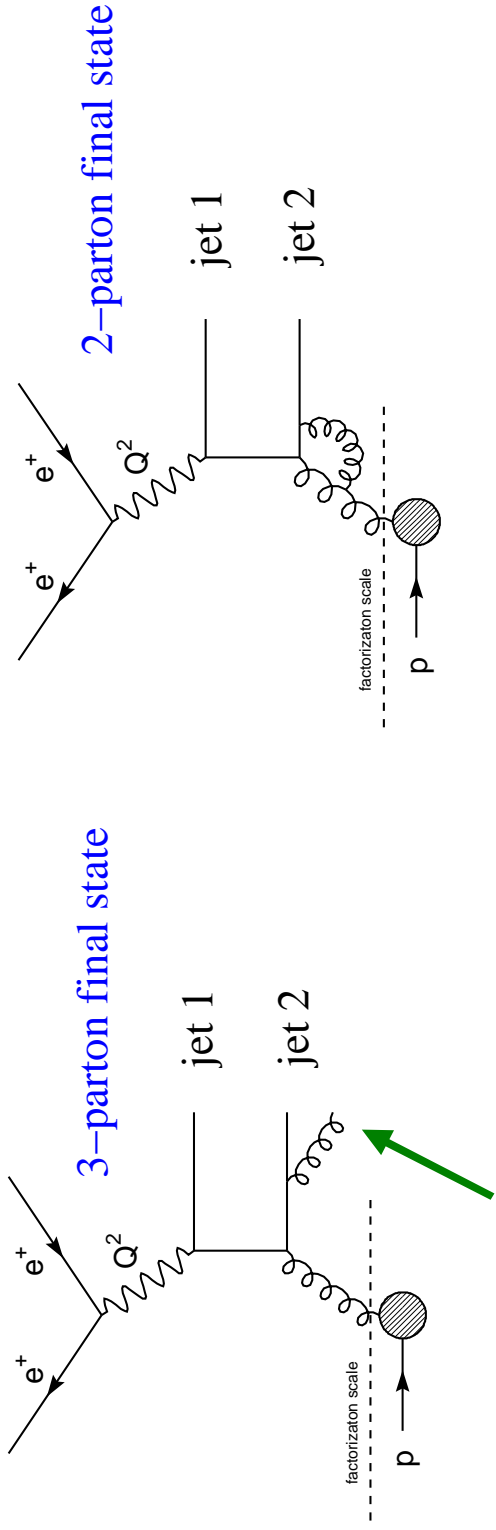

Soft gluon emission reduces transverse energy of second jet

Symmetric cut ($E_{T,jet\,1} = E_{T,jet\,2}$) limits the 3-body phase space, introducing infrared sensitivity and disrupting the compensation between real and virtual corrections.

Asymmetric cut $\Rightarrow E_{T,jet\,2}\,(\text{cut}) < E_{T,jet\,1}$

S. Frixione and G. Ridolfi
Nucl.Phys.B 507 (1997)

Recontres de Moriond – 3/27/02    Dijet Measurements at ZEUS    3

# Asymmetric Jet Cuts II: Study with DISENT

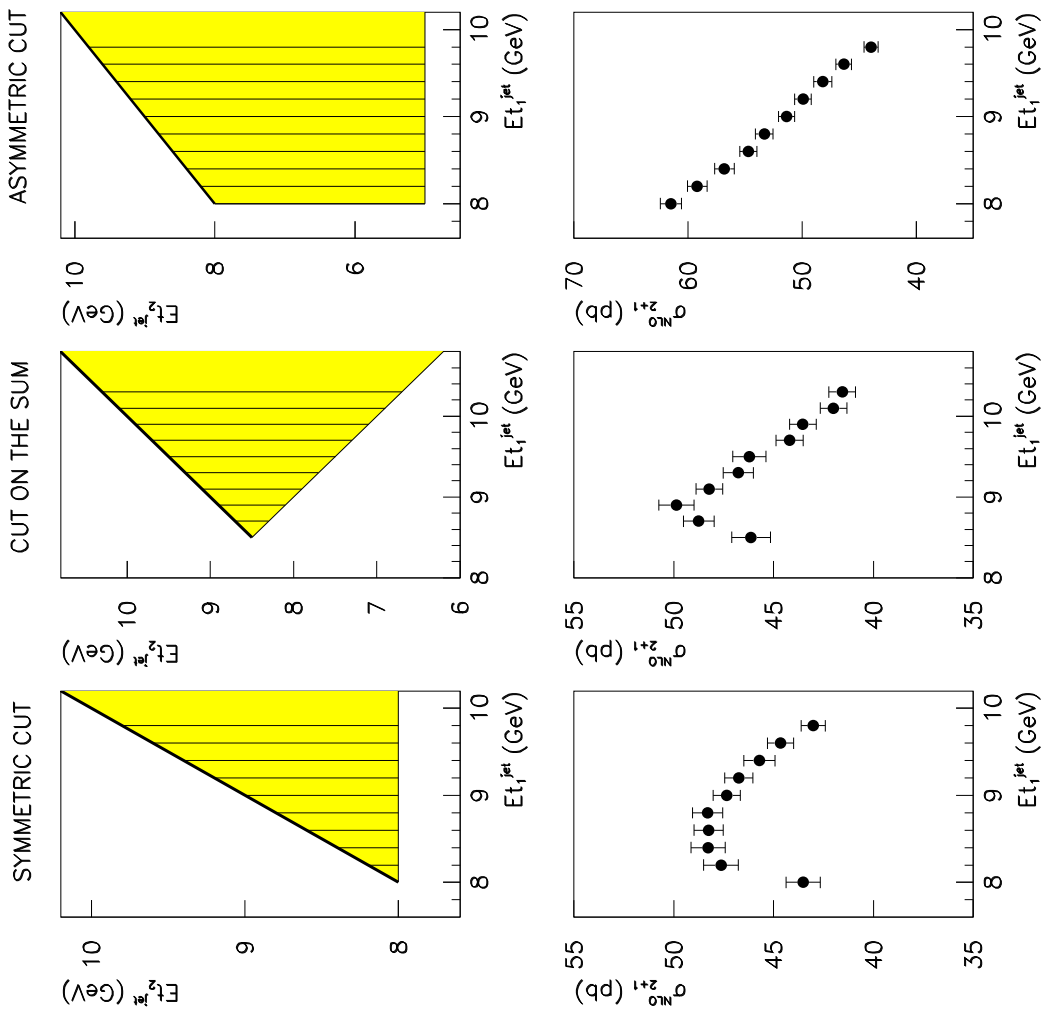

To study the effect, 3 scenarios were considered:

(a) $E_{T,\text{jet }1} > 8$ GeV
$E_{T,\text{jet }2} > 8$ GeV

(b) $E_{T,\text{jet }1} > 5$ GeV
$E_{T,\text{jet }2} > 5$ GeV
$E_{T,\text{jet }1} + E_{T,\text{jet }2} > 17$ GeV

(c) $E_{T,\text{jet }1} > 8$ GeV
$E_{T,\text{jet }2} > 5$ GeV

Purely asymmetric jet cuts cure dijet cross section of unphysical behavior near the symmetric cut

# Impact on Dijet Analyses

*With asymmetric cuts, agreement between dijet measured cross sections and NLO predictions in a variety of different observables!*

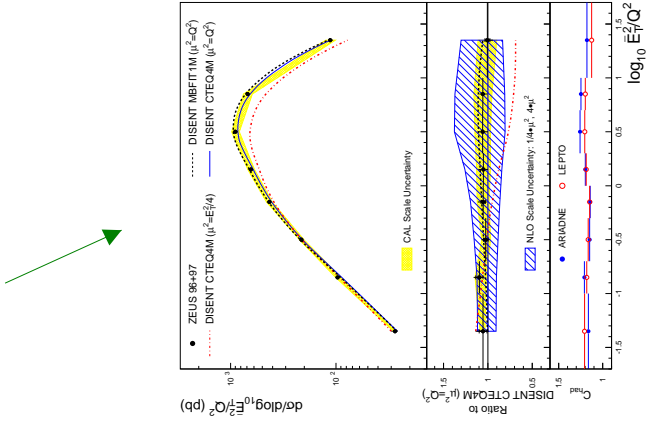

## Data and NLO comparison: symmetric vs. asymmetric cuts

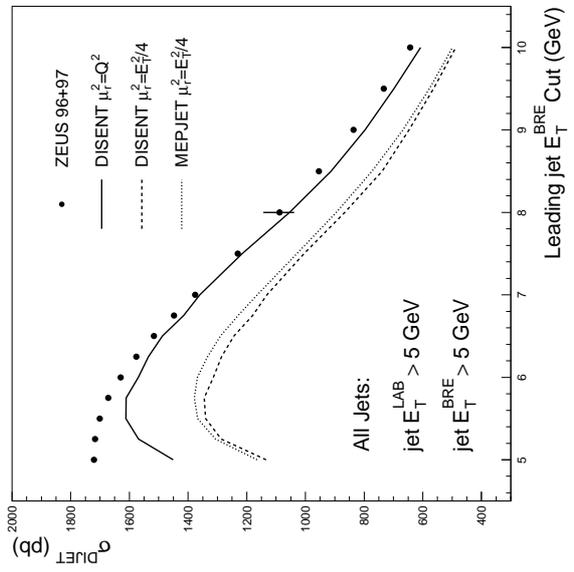

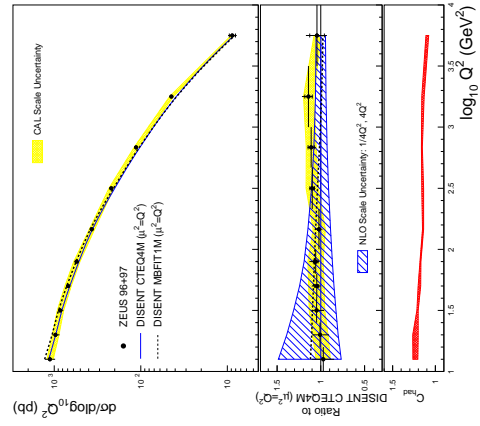

**Conclusion:**
Asymmetric jet cuts give increased phase space region for 3-parton final states, allowing proper cancellation of NLO corrections and physically meaningful cross sections.

Infrared sensitivity observed with other NLO programs:
MEPJET : see hep-ex/0109029
JetVip : see B.Potter, Comput.Phys.Commun. 133(2000)